\documentclass[12pt,letter]{article}
\begin{document}
\thispagestyle{empty}

\mbox{}
%\vspace{0.5cm}
\begin{center}
{\large{\bf THE EINSTEIN-BOLTZMANN RELATION FOR THERMODYNAMIC AND HYDRODYNAMIC FLUCTUATIONS}}

\vspace{0.5cm}

{\it A. J. McKane$^{\rm a}$, F. V\'azquez$^{\rm b}$ and M. A.
Olivares-Robles$^{\rm b,\, c}$}
\\
\bigskip

$^{\rm\, a}$Theory Group, School of Physics and Astronomy, \\
University of Manchester, Manchester M13 9PL, UK

\medskip
$^{\rm\, b}$Facultad de Ciencias, Universidad Aut\'{o}noma del
Estado de
Morelos, \\
Avenida Universidad 1001, Chamilpa, Cuernavaca, Morelos 62209, M\'{e}xico \\

\medskip
$^{\rm\, c}$Secci\'{o}n de Posgrado e Investigaci\'{o}n, Escuela Superior
de Ingenier\'{\i}a Mec\'{a}nica y El\'{e}ctrica Culhuacan-IPN, Av. Santa
Ana 1000, Col. San Francisco Culhuacan Coyoacan 04430, M\'{e}xico D.F.\\
\end{center}

\vspace{2cm}

\begin{abstract}
When making the connection between the thermodynamics of
irreversible processes and the theory of stochastic processes
through the fluctuation-dissipation theorem, it is necessary to
invoke a postulate of the Einstein-Boltzmann type. For convective
processes hydrodynamic fluctuations must be included, the velocity
is a dynamical variable and although the entropy cannot depend
directly on the velocity, $\delta^{2} S$ will depend on velocity
variations. Some authors do not include velocity variations in
$\delta^{2} S$, and so have to introduce a non-thermodynamic
function which replaces the entropy and does depend on the
velocity.  At first sight, it seems that the introduction of such
a function requires a generalisation of the Einstein-Boltzmann
relation to be invoked. We review the reason why it is not
necessary to introduce such a function, and therefore why there is
no need to generalise the Einstein-Boltzmann relation in this way.
We then obtain the fluctuation-dissipation theorem which shows
some differences as compared with the non-convective case. We also
show that $\delta^{2} S$ is a Liapunov function when it includes
velocity fluctuations.
\end{abstract}

\vspace{0.5cm}

\newpage

\section{\large INTRODUCTION}
Velocity fluctuations play an important role in a variety of non-equilibrium
phenomena. Mention can be made, for instance, of time dependent diffusion
processes in binary liquid mixtures, where they are the principal mechanism
leading to anomalously large fluctuations in concentration \cite{vailati}.
Also, the coupling between temperature and transverse-velocity fluctuations in
the well known case of a horizontal fluid layer heated from below may be
associated with a small convective heat transfer below the Rayleigh-B\'{e}nard
instability \cite{ortiz}. It is natural to consider these kind of problems
from the point of view of irreversible thermodynamics. However, there is no
prescription for how to introduce the velocity fluctuations into the formalism.

The standard method of introducing fluctuations into irreversible
thermodynamics is through the Einstein-Boltzmann relation
$P_{S} \sim \exp\left\{ \delta^{2} S/2k_{B}\right\}$, where $P_{S}$ is the
stationary probability distribution and $\delta^{2} S$ is the second
variation of the local entropy \cite{deg84}. In this paper we will be
interested in convective processes where the velocity is included as a
dynamical variable, and in the explicit form for $\delta^{2} S$
in this case. It should be noted, and is widely appreciated,
that the entropy does not depend directly on the velocity of the
system: velocity is a hydrodynamic, but not a thermodynamic variable.
Therefore some authors, notably Glansdorff and Prigogine \cite{gla71},
do not include velocity variations in the expression for $\delta^{2} S$.

A solution to this problem could be to introduce a new function which
is essentially a generalisation of the entropy which does depend on
the velocity. This would not be a thermodynamic function, but it
would then be necessary to generalise the Einstein-Boltzmann
relation in such a way that entropy would be replaced by this new
function. Such a function has been introduced some time
ago by Glansdorff and Prigogine, but in the context of thermodynamic
and hydrodynamic stability \cite{gla71}. They suggested defining a
new function $z \equiv s - \mathbf{v}^{2}/2T_{0}$, where $s$ is the
entropy per unit mass, $\mathbf{v}$ is the barycentric velocity and
$T_{0}$ is the temperature in the reference state (for example, the
temperature in equilibrium).  The analogous quantity for the system
as a whole will be denoted by $Z$ and is given by $Z = \int \rho z dV$,
just as $S = \int \rho s dV$. This function has not been utilised a great
deal, perhaps in part because among those who explicitly use the
$Z$-function \cite{gla71}-\cite{mat76}, most do not consistently use
the definition given above, sometimes using the (varying)
temperature $T$ in place of the (non-varying) reference temperature $T_{0}$.

The main reason why the function $Z$ has not been widely used is
no doubt the demonstration by Oono \cite{oon76} that $\delta^{2} S$
does in fact contain velocity variations, even though the entropy
does not depend on the velocity. In fact, the entropy may be
written in terms of the velocity if other variables are introduced
which exactly cancel out the velocity dependence \cite{lan80}. To see this
let us write \cite{oon76}
\begin{equation}
dU = TdS-pdV+\mu_{\gamma}dN_{\gamma},
\label{gibbs}
\end{equation}
where $U$ is the internal energy, $V$ the volume, $N_{\gamma}$ the number
of moles of the $\gamma$-chemical species, $p$ the pressure, and
$\mu_{\gamma}$ the chemical potential of the $\gamma$-species. In addition,
let $E_T$ be the total energy:
\begin{equation}
E_T=U+mv_{\mu}v_{\mu}/2,\label{Etotal}
\end{equation}
$v_{\mu}$ being the barycentric velocity and $m$ the mass. We assume that the
changes in potential energy due to altitude, for instance, are negligible.
Therefore we omit a potential term in this definition. Then Eq. (\ref{gibbs})
can be written as
\begin{equation}
dE_T = TdS-pdV+\mu_{\gamma}dN_{\gamma}+mv_{\mu}dv_{\mu}.
\label{gibbs1}
\end{equation}
Now note that in Eq. (\ref{gibbs1}) the term $dE_T-mv_{\mu}dv_{\mu}$ does not
depend on velocity in accordance with the definition of the total energy
Eq.(\ref{Etotal}). So the entropy in Eq. (\ref{gibbs1}) does not depend on
the velocity and the thermodynamic consistency of this form of Gibbs relation
Eq.(\ref{gibbs}) is ensured.

Oono also showed that $\delta^{2} Z$ is nothing else but $\delta^{2} S$.
However, mention must be made of the fact that $\delta^{2} S$ and
$\delta^{2} Z$ are only equal within approximation schemes where $T$ can be
replaced by $T_{0}$. There is also a lack of consensus as to whether
$\delta^{2} S$ is a Liapunov function in systems where velocity is a
dynamical variable: some authors believe it is \cite{fox70}, others
believe it is not \cite{gla71}. Some of this confusion involves matters
of principle, some involves matters of notation (for instance,
$\delta^{2} S$ meaning two entirely different things), and some involves
inconsistencies in definitions of key quantities. Our objective in this
paper is to clarify many of these points, by examining their consequences
in the context of linear theories of irreversible thermodynamics, and to
obtain the explicit form of the fluctuation-dissipation theorem for
convective processes. We remark in passing that there are a whole
set of different subtleties and controversies in extending these ideas to
the non-linear regime \cite{lan72,lav72}, but we do not explore these here.
\section{\large IRREVERSIBLE THERMODYNAMICS AND\\
STOCHASTIC PROCESSES}
A fluid being described within linear irreversible thermodynamics (LIT)
requires five local variables: the volume per unit mass $v$, the barycentric
velocity $v_{\mu}$ and the temperature $T$ \cite{deg84}, but our conclusions
will be more widely applicable, for example applying also to a fluid in
extended irreversible thermodynamics (EIT), which requires 14
dynamic variables \cite{jou96}. To keep the notation general, we
will denote the fluctuations in the independent  dynamic variables
as $a_{b} (\mathbf{r}, t)$, where $b=1,\ldots,N$ and assume that
they satisfy a set of Langevin-type equations:
\begin{equation}
\frac{\partial a_{b} (\mathbf{r}, t)}{\partial t} = - \sum_{c}\,\int\,
d\mathbf{r}^{\prime} G_{b c} (\mathbf{r}, \mathbf{r}^{\prime})
a_{c} (\mathbf{r}^{\prime}, t) + \tilde{f}_{b} (\mathbf{r}, t)\,.
\label{langevin}
\end{equation}
Here the first term on the right-hand side is a result of the linearization
of the macroscopic equation about the stationary state and
$\tilde{f}_{b} (\mathbf{r}, t)$ is a stochastic term that represents
fluctuations in the system. For the particular case of a fluid within LIT $N=5$
and the five local variables $a_{1},\ldots,a_{5}$ are the scaled versions of
fluctuations in $\{ v, v_{\mu}, T\}$. Specifically if the equilibrium state is
denoted by $\{ v_{0}, 0, T_{0}\}$, and fluctuations away from this state by
$\{ v_{1}, v_{\mu}, T_{1}\}$, then we define the $a_{b}$ by
\cite{fox70, mck01}:
\begin{equation}
a_{1}=-\rho _{0}^{\frac{3}{2}}v_{1}\ ,
\ a_{\mu + 1 }=\frac{\rho _{0}^{1/2}}{c_T} v_{\mu}\ , \
a_{5}=\left( \frac{\rho _{0}C_v}{T_{0} c_{T}^{2}}\right)^{\frac{1}{2}}T_{1}\,,
\label{new}
\end{equation}
with $\mu=1,2,3$. Here $\rho_0$ is the mass density, $c_{T}$ the isothermal
speed of sound and $C_v$ the specific heat at constant volume, all in
equilibrium. These rescalings simplify the algebraic structure of the results.
We use the same notation for the velocity and the velocity fluctuations,
since no confusion should arise.

The analysis of the fluctuations is made more transparent if we adopt an
abbreviated form where the continuous labels $\mathbf{r}$ and
$\mathbf{r}^{\prime}$ are replaced by the discrete labels $j$ and $k$ and where
the summation convention is assumed. In this case, (\ref{langevin}) becomes
\begin{equation}
\dot{a}_b^j(t)+G_{bc}^{jk}a_c^k(t)=\tilde{f}_b^j(t)\ ;\ \ b,c=1,...,N\,.
\label{lang}
\end{equation}
To complete the specification of the stochastic dynamics, the statistics of
the stochastic terms $\tilde{f}_b^j(t)$ need to be given. We will take them
to have a Gaussian distribution with mean zero and correlator
\begin{equation}
\left\langle \tilde{f}_b^j(t)\,\tilde{f}_c^k(t^{\prime })\right\rangle
=2Q_{bc}^{jk}\delta (t-t^{\prime })\,.
\label{correlator}
\end{equation}
The requirement that they have zero mean follows from the fact that we ask
that the $a_b$ have zero mean: $\langle a^{j}_{b} \rangle = 0$. The matrix
$Q$ is real, symmetric and positive semidefinite. We will not give an
explicit form for the matrix $G$ here: it may be straightforwardly derived
by a linearization of the macroscopic equations \cite{fox70}. As will be
discussed below, the matrix $Q$ may be given in terms of the matrix $G$ and
another matrix $E$, which is the covariant matrix of the $a^{k}_{b}$ in the
stationary state:
\begin{equation}
\langle a_e^la_f^m\rangle _S=(E^{-1})_{ef}^{lm}\,.
\label{stat_corr}
\end{equation}
Therefore the stochastic dynamics will be completely specified if we can
determine the matrix $E$. Clearly we need some new information from
which to find it. This is the Einstein-Boltzmann relation.

The Gaussian assumption determines the class of phenomena to be dealt with. In
general, the Gaussian assumption is valid for a wide range of conditions
in which the physical variables do not change too fast with time \cite{deg84}.
It may be said that the sufficient condition for the validity of this
assumption is the local equilibrium hypothesis. Nevertheless, the system may
be in a non-equilibrium non-stationary state in which such a hypothesis is
not satisfied and yet will be well described throughout using the Gaussian
assumption.

We now introduce the fluctuation-dissipation theorem by recalling
that another way of specifying the stochastic process defined by
Eqs. (\ref{lang}) and (\ref{correlator}) is through the
Fokker-Planck equation \cite{gar85,ris89}
\begin{equation}
\frac{\partial P(\underline{\mathbf{a}}, t)}{\partial t} =
\frac{\partial }{\partial a^{j}_{b}} \left[ G^{jk}_{bc} a^{k}_{c}
P(\underline{\mathbf{a}}, t) \right] + \frac{\partial^{2} }
{\partial a^{j}_{b} \partial a^{k}_{c}} \left[ Q^{jk}_{bc}
P(\underline{\mathbf{a}}, t) \right], \label{FPE}
\end{equation}
where $P(\underline{\mathbf{a}}, t)$ is the probability distribution function
of the local variables $\mathbf{a}$. This is a linear Fokker-Planck equation
and so the solution is a Gaussian which may be written down explicitly
as \cite{van92}
\begin{eqnarray}
P(\underline{\mathbf{a}}, t) &=& \mathcal{N}\,
\left( \det \Xi (t) \right)^{-1/2} \nonumber \\
&\times& \exp\left\{ - \frac{1}{2} \underline{\mathbf{a}}^{T} \Xi
(t)^{-1} \underline{\mathbf{a}} \right\}\,, \label{FP_soln}
\end{eqnarray}
where $\mathcal{N}$ is a normalisation constant and where the
matrix $\Xi (t)$ is given by
\begin{equation}
\Xi (t-t_{0}) = 2 \int^{t}_{t_{0}} e^{-\left( t-t^{\prime}
\right)G}\,Q\, e^{\left( t-t^{\prime} \right)G}\,dt^{\prime}\,.
\label{xi}
\end{equation}
Here initial conditions have been set at $t=t_{0}$ and we have
made use of the fact that $\langle a^{j}_{b} \rangle=0$. By
letting $t_{0} \to -\infty$, we find the stationary distribution.
It has the form (\ref{FP_soln}), but with $\Xi (t)$ replaced by
\begin{eqnarray}
\Xi (\infty) &=& 2 \int^{t}_{-\infty} e^{-\left( t-t^{\prime}
\right)G}\,Q\,
e^{\left( t-t^{\prime} \right)G}\,dt^{\prime} \nonumber \\
&=& 2 \int^{\infty}_{0} e^{- \rho G}\,Q\, e^{ \rho G}\,d\rho\,.
\label{xi_stat}
\end{eqnarray}

To make use of the Einstein-Boltzmann relation, let us observe that since
the $a^{k}_{b}$ have zero mean, and since they are linearly related to
the $f^{k}_{b}$ which are Gaussian, they also have a Gaussian distribution
with a stationary probability distribution of the form:
\begin{equation}
P_S(\underline{\mathbf{a}})=\mathcal{N}\,\exp \left\{ -\frac
12a_b^jE_{bc}^{jk}a_c^k\right\} \,.
\label{stat_prob}
\end{equation}
Here $\underline{\mathbf{a}}=(\underline{a}^{1}, \underline{a}^{2},\ldots)$
where $\underline{a}^{i}=(a^{i}_{1},\ldots,a^{i}_{N})$ and $\mathcal{N}$ is
a normalisation constant. By comparing (\ref{stat_prob}) with (\ref{FP_soln})
when $t_{0} \to -\infty$, we can make the identification
\begin{equation}
E^{-1} = \Xi (\infty) = 2 \int^{\infty}_{0} e^{- \rho G}\,Q\, e^{
\rho G}\,d\rho\,. \label{identification}
\end{equation}
Performing the integral in (\ref{identification}) gives the result \cite{van92}
\begin{equation}
2 Q^{ij}_{ab} = G^{ik}_{ac} (E^{-1})^{kj}_{cb} +
(E^{-1})^{ik}_{ac} G^{T\,kj}_{cb}\,, \label{fdt}
\end{equation}
where T denotes transpose. This is the fluctuation-dissipation
theorem of the theory. It is the required relationship which gives
the matrix $Q$ in terms of the matrices $G$ and $E$.
\section{\large THE FLUCTUATION-DISSIPATION THEOREM FOR CONVECTIVE SYSTEMS}
The result (\ref{stat_prob}) may be
compared directly \cite{mck01} with the Einstein-Boltzmann relation
\begin{equation}
P_{S}(\underline{\mathbf{a}}) \sim \exp\left\{\delta^{2}S/2k_{B} \right\}\,,
\label{einstein}
\end{equation}
so that
\begin{equation}
S(\underline{\mathbf{a}}) = S_{\rm eq} - \frac{1}{2} k_{B}
a_b^j E_{bc}^{jk}a_c^k\,.
\label{stat_S}
\end{equation}
The indices $b$ and $c$ in (\ref{stat_prob}) or (\ref{stat_S}) run
from $1$ to $N$ (from $1$ to $5$ in LIT) and include the velocity
as a variable. However, if only specific volume (or density) and
temperature are included as variables in $\delta^{2} S$
\cite{gla71,cal60}, then it apparently seems that Eqs. (\ref{stat_prob}) and
(\ref{einstein}) cannot be compared to determine the $E^{jk}_{bc}$
matrix. Thus, it seems clear that the $\delta^{2} S$ which we need to use in
the Einstein-Boltzmann relation is the one which allows for variations in the
velocity. In fact, as shown by Oono \cite{oon76},
\begin{eqnarray}
\delta^{2} S  &=& \delta \left( \frac{1}{T} \right) \delta U +
\delta \left( \frac{p}{T} \right) \delta V - \frac{m\delta v_{\mu}
\delta v_{\mu}}{T} \nonumber \\
&=& \left. \delta^{2} S \right|_{\bf v} - \frac{m\delta v_{\mu}
\delta v_{\mu}}{T}\,,
\label{oono_relation}
\end{eqnarray}
where $\delta^{2} S|_{\bf v}$ is $\delta^{2} S$ with no variation in the
velocity. Using $\delta^{2} S$, rather than $\delta^{2} S|_{\bf v}$ allows
Eqs. (\ref{stat_prob}) and (\ref{einstein}) to be compared and the matrix $E$
determined. It should be noted that (i) in Ref. \cite{mck01} the additional
term to be added to $\delta^{2} S|_{\bf v}$ was given as
$m\delta v_{\mu} \delta ( - v_{\mu}/T)$, and (ii) in Ref. \cite{oon76} it was
stated that $\delta^{2} Z = \delta^{2} S$ --- whereas from the definition of
$z$ we see that
\begin{equation}
\delta^{2} Z  = \left. \delta^{2} S \right|_{\bf v} - \frac{m\delta v_{\mu}
\delta v_{\mu}}{T_{0}}\,.
\label{oono_relation2}
\end{equation}
Both the results (i) and (ii) are true in the linear regime, where
$T^{-1}$ may be replaced by $T^{-1}_{0}$, but they are not true in general;
the correct form for $\delta^{2} S$ is given in Eq.
(\ref{oono_relation}), and $\delta^{2} Z$ is not equal to
$\delta^{2} S$, it is given by Eq. (\ref{oono_relation2}). A
consequence of this is that in the linear regime the
Einstein-Boltzmann relation may also be written as $P_{S} \sim
\exp\left\{ \delta^{2} Z/2k_{B}\right\}$. This means that if we
were to use $\delta^{2} S|_{\bf v}$, as Glansdorff and Prigogine
do, we would need to invoke this latter form of the
Einstein-Boltzmann relation to identify the matrix $E$ and so make
the connection between irreversible thermodynamics and the theory
of stochastic processes, at least in the linear regime. However,
as we have stressed there is no need to introduce this extra
postulate, and we may use the usual form $P_{S} \sim \exp\left\{
\delta^{2} S/2k_{B}\right\}$, as long as the correct form of
$\delta^{2} S$ (\ref{oono_relation}) is used.

We can now come back to the task of determining the matrix $E$.
Let us first write down the expression for $\delta^{2} S$ without
velocity variations in terms of the scaled versions of $v_1$ and
$T_1$, namely $a_1$ and $a_5$ to see explicitly where the process
fails. After some straightforward manipulations \cite{mck01} of this
standard result \cite{cal60}, we obtain, using the
Einstein-Boltzmann relation,
\begin{equation}
P_S(\underline{\mathbf{a}}) \sim \exp\left\{ \frac{c_{T}^{2}}
{2 k_{B} T_{0}} \left[ - a^{j}_{1} a^{j}_{1}
- a^{j}_{5} a^{j}_{5} \right] \right\}\,.
\label{EBresult}
\end{equation}
If this result were to be compared with (\ref{stat_prob}) then it would
imply that $E$ would be diagonal, but with entries corresponding
to the velocity fluctuations being zero. This is clearly not
correct since, for instance, the velocity-velocity correlation
function in equilibrium (\ref{stat_corr}) would be formally
infinite. Using instead the form of $\delta^{2} S$ allowing
for velocity variation we find
\begin{equation}
P_S(\underline{\mathbf{a}}) \sim \exp\left\{ \frac{c_{T}^{2}}
{2 k_{B} T_{0}} \left[ - a^{j}_{b} a^{j}_{b} \right] \right\}\,,
\label{EBPresult}
\end{equation}
since $v_{\mu}=(c_{T}^{2}/\rho_{0})^{1/2} a_{\mu + 1}$ and where $b=1,...,5$.
A comparison with (\ref{stat_corr}) gives the identification
\begin{equation}
E^{jk}_{bc} = \frac{c_{T}^{2}}{k_{B} T_{0}}\,\delta_{jk}\,
\delta_{bc}\,.
\label{E}
\end{equation}
This now gives a consistent result, which when used in conjunction
with the fluctuation-dissipation theorem (\ref{fdt}), completely
specifies the stochastic dynamics described by (\ref{lang}) and
(\ref{correlator}) or by (\ref{FPE}). An explicit expression for
matrix $Q$ is obtained by substituting Eq. (\ref{E}) into Eq.
(\ref{fdt}). The result is
\begin{equation}
Q_{bc}^{jk}=\frac{k_BT_0}{A}\,S_{bc}^{jk}\ .  \label{Q_S_abbrev}
\end{equation}
where $S_{bc}^{jk}$ represents the symmetric part of the dynamic
matrix $G$:
\begin{eqnarray}
S_{\mu+1, \nu+1}(\mathbf{r}, \mathbf{r}^{\prime}) & = & \frac{1}{\rho_{0}}%
\,\left[ 2\mu X_{\mu \rho \nu \sigma} + \zeta \delta_{\mu \rho} \delta_{\nu
\sigma} \right]\, \frac{\partial^{2}\,}{\partial x_{\rho} \partial
x_{\sigma}^{\prime}} \delta(\mathbf{r} - \mathbf{r}^{\prime})\ ,
\label{first_S_LIT} \\
S_{5 5}(\mathbf{r}, \mathbf{r}^{\prime}) & = & \frac{1}{\rho_{0}C}\, \lambda
\delta_{\mu \nu}\, \frac{\partial^{2}\,}{\partial x_{\mu} \partial
x_{\nu}^{\prime}}\delta(\mathbf{r} - \mathbf{r}^{\prime})\ ,
\label{second_S_LIT}
\end{eqnarray}
with all other $S_{bc}(\mathbf{r}, \mathbf{r'})$, including
$S_{11}(\mathbf{r}, \mathbf{r'})$, equal to zero. The tensor
$X_{\mu\nu\rho\sigma}$ is defined by
\begin{equation}
X_{\mu \nu \rho \sigma} = \frac{1}{2} \left( \delta_{\mu \rho} \delta_{\nu
\sigma} + \delta_{\mu \sigma} \delta_{\nu \rho} - \frac{2}{3}\,\delta_{\mu
\nu} \delta_{\rho \sigma} \right)\ .  \label{X}
\end{equation}
In Eqs. (\ref{first_S_LIT}) and (\ref{second_S_LIT}), the continuum
limit has been taken so that the discrete spatial variables $j,k$
have been replaced by $\mathbf{r}, \mathbf{r'}$. As mentioned
above, all the matrices in Eq.(\ref{fdt}) are $5\times 5$ in the
convective case, unlike in the non-convective case where they
are $2\times 2$.

The discussion above took place within the framework of LIT
which contains 5 dynamical variables, but the idea is more
general. We have already mentioned EIT where the dissipative
fluxes are raised to the same status as the thermodynamic
variables. In this case $\delta^{2} S$ (where $S$ now denotes the
corresponding non-equilibrium thermodynamic potential in place of
the local equilibrium entropy) contains terms involving
these fluxes, as well as the more conventional thermodynamical
variables, but not the velocity variables \cite{jou96}. Written in
terms of scaled variables it has the form \cite{mck01}
\begin{displaymath}
P_S(\underline{\mathbf{a}}) \sim \exp\left\{ \frac{c_{T}^{2}}
{2 k_{B} T_{0}} \left[ - a^{j}_{1} a^{j}_{1}
- a^{j}_{5} a^{j}_{5} - \frac{1}{2} \stackrel{o}{a}^{j}_{\mu \nu}
\stackrel{o}{a}^{j}_{\nu \mu} \right. \right.
\end{displaymath}
\begin{equation}
\left. \left. - a^{j}_{\mu + 10} a^{j}_{\mu + 10} - a^{j}_{14} a^{j}_{14}
\right] \right\}\,.
\label{EITresult}
\end{equation}
Here the variables $\stackrel{o}{a}^{j}_{\mu \nu}, a^{j}_{\mu + 10}$
and $a^{j}_{14}$ are scaled versions of the traceless stress tensor,
the heat flux and the trace of the stress tensor, respectively.
The result (\ref{EITresult}) suffers from the same defect as
(\ref{EBresult}), but if we now include the velocity variations in
$\delta^{2} S$ then  we again obtain (\ref{EBPresult}), but now with
$b=1,\ldots,14$. Therefore the matrix $E$ can be consistently
identified, and again is given by (\ref{E}).
\section{\large VELOCITY FLUCTUATIONS AND THE\\ LIAPUNOV FUNCTION}
Finally, within the context of LIT or EIT, we can investigate the claim
that $\delta^{2} Z$ is a Liapunov function, but that $\delta^{2} S$ can
no longer be adopted as a Liapunov function when velocity is included
as a dynamical variable \cite{gla71}.  In the language we have been
using in this paper, the former is $\delta^{2} S$ and the latter is
$\delta^{2} S|_{\bf v}$, and this is the notation we will use in what
follows. To investigate whether these functions are Liapunov
functions, we begin from the form of $\delta^{2} S$ sufficiently near
equilibrium that LIT will apply:
\begin{equation}
\delta^{2} S = - \frac{c_{T}^{2}}{T_{0}} a^{j}_{b} (t) a^{j}_{b} (t)\,.
\label{Z_for_stability}
\end{equation}
Here the $a^{j}_{b}$ are averaged variables, that is, non-fluctuating
variables which obey the hydrodynamic balance equations. From
Eq. (\ref{Z_for_stability}) we see that $\delta^{2} S \leq 0$ with
equality if and only if $a^{j}_{b} (t)= 0$. Differentiating Eq.
(\ref{Z_for_stability}) with respect to time gives
\begin{eqnarray}
\frac{d }{dt} \left( \delta^{2} S \right) = - \frac{2 c_{T}^{2}}{T_{0}}
\dot{a}^{j}_{b} (t) a^{j}_{b} (t) &=& \frac{2 c_{T}^{2}}{T_{0}}
G_{bc}^{jk}a_c^k(t) a_{b}^{j} (t) \nonumber \\
&=& \frac{2 c_{T}^{2}}{T_{0}}
S_{bc}^{jk}a_c^k(t) a_{b}^{j} (t)\,,
\label{Z_dot}
\end{eqnarray}
where $S_{bc}^{jk}$ is the symmetric part of $G_{bc}^{jk}$. Using
the expressions for $S_{bc}^{jk}$, Eqs. (\ref{first_S_LIT}) and
(\ref{second_S_LIT}), and integrating by parts gives
\begin{equation}
\frac{d }{dt} \left( \delta^{2} S \right) = \frac{2 c_{T}^{2}}{\rho_{0} T_{0}}
\int d\mathbf{r}\,\left( 2\mu \stackrel{\circ}{D}_{\mu \nu}
\stackrel{\circ}{D}_{\mu \nu} + \zeta D_{\mu \mu}^{2} +
\frac{\lambda}{C_{v}} \frac{\partial a_{5}}{\partial x_{\mu}}
\frac{\partial a_{5}}{\partial x_{\mu}} \right) \geq 0\,,
\label{Lia_Z_LIT}
\end{equation}
where we have gone back to an explicit notation for the continuous
space variable $\mathbf{r}$. In Eq. (\ref{Lia_Z_LIT}), $\lambda, \zeta$
and $\mu$ are the thermal conductivity, the bulk viscosity and the
shear viscosity, respectively, $D_{\mu \nu}$ is the symmetric
part of the scaled velocity gradient and $\stackrel{\circ}{D}_{\mu \nu}$
its traceless form:
\begin{equation}
D_{\mu \nu} = \frac{1}{2} \left( \frac{\partial a_{\mu + 1}}{\partial x_{\nu}}
+ \frac{\partial a_{\nu + 1}}{\partial x_{\mu}} \right)\,, \ \
\stackrel{\circ}{D}_{\mu \nu} = D_{\mu \nu} - \frac{1}{3} D_{\rho \rho}\,
\delta_{\mu \nu}\,.
\label{defn_Ds}
\end{equation}
This shows explicitly, when $\delta^{2} S$ is defined in terms of
the averaged variables, that it is a Liapunov function, as
suggested by Glansdorff and Prigogine \cite{gla71}. However, this
calculation is identical to one carried out in Ref. \cite{fox70}, where
$dS/dt$ was evaluated and shown to be non-negative. Since all of
these calculations have been carried out in the linear regime, and
$dS/dt = (1/2) d( \delta^{2} S)/dt = (1/2) d( \delta^{2} Z)/dt$ this is not
surprising. Note that the inequality in Eq. (\ref{Lia_Z_LIT}) is an
equality if and only if
$\stackrel{\circ}{D}_{\mu \nu} = 0, D_{\mu \mu} = 0$ and
$\partial a_{5}/\partial x_{\mu} = 0$. From the constitutive relations
for LIT this corresponds to the vanishing of the traceless stress tensor
and its trace and of the heat flux. This condition corresponds to the
thermodynamic equilibrium state and it is equivalent to the condition
$a_{b}^{j} (t) = 0$ found when $\delta^{2} S$ given by Eq.
(\ref{Z_for_stability}) is equal to zero.

A similar calculation may be carried out for EIT. In this case
Eqs. (\ref{Z_for_stability}) and (\ref{Z_dot}) also hold, but now
with the indices $b$ and $c$ running from 1 to 14. The form of the
$S_{bc}^{jk}$ are different for EIT --- in some ways they are
simpler, since they do not involve derivatives, and so no
integration by parts is required to obtain an explicit expression
for the time derivative of $\delta^{2} S$. Using the expressions
for $S_{bc}^{jk}$ given in Ref. \cite{mck01} for EIT one finds
that
\begin{equation}
\frac{d }{dt} \left( \delta^{2} S \right) = \frac{2 c_{T}^{2}}{T_{0}}
\int d\mathbf{r}\,\left( \frac{1}{2} \tau_{2}^{-1}
\stackrel{\circ}{a}_{\mu \nu}
\stackrel{\circ}{a}_{\nu \mu} + \tau_{0}^{-1} a_{14} a_{14} +
\tau_{1}^{-1} a_{\mu + 10} a_{\mu + 10} \right) \geq 0\,,
\label{Lia_Z_EIT}
\end{equation}
where the $\tau_{i},\ i=0,1,2$ are the relaxation times of the various
fluxes. Once again $\delta^{2} S$ is seen to be a Liapunov function,
with the inequality in Eq. (\ref{Lia_Z_EIT}) becoming an equality if
and only if $\stackrel{\circ}{a}_{\mu \nu} = 0, a_{14} = 0$ and
$a_{\mu + 10} = 0$. These are just scaled versions of the traceless
stress tensor and its trace, and of the heat flux, and so equality is
obtained when these vanish, just as for LIT. If we use this method
to try and show that $\delta^{2} S|_{\bf v}$ is a Liapunov function, we
find, for example in the case of LIT,
\begin{equation}
\left. \delta^{2} S \right|_{\mathbf{v}} = - \frac{c_{T}^{2}}{T_{0}}
\left( a^{j}_{1} (t) a^{j}_{1} (t) + a^{j}_{5} (t) a^{j}_{5} (t) \right)\,,
\label{S_for_stability}
\end{equation}
and differentiating with respect to time gives
\begin{eqnarray}
\frac{d }{dt} \left( \left. \delta^{2} S\right|_{\mathbf{v}} \right) &=&
- \frac{2 c_{T}^{2}}{T_{0}} \left( \dot{a}^{j}_{1} (t) a^{j}_{1} (t) +
\dot{a}^{j}_{5} (t) a^{j}_{5} (t) \right) \nonumber \\
&=& \frac{2 c_{T}^{2}}{T_{0}}
\left( G_{1c}^{jk}a_c^k(t) a_{1}^{j} (t) +
G_{5c}^{jk}a_c^k(t) a_{5}^{j} (t) \right)\,.
\label{S_dot}
\end{eqnarray}
Substituting the actual expressions for $G_{bc}^{jk}$ \cite{fox70,mck01}
in Eq. (\ref{S_dot}), does not give an expression which is manifestly
positive semi-definite. This is no doubt what Glansdorff and Prigogine
meant by saying that $\delta^{2} S$ loses its properties as a Liapunov
function when velocity is included as a dynamical variable. However, since
we are assuming that ${\bf v}$ is fixed in the definition of $\delta^{2} S$ it
might be more consistent to take $\mathbf{v}$ to be a constant in the
balance equations. If we do this we find that only the third term in the
parentheses in Eq. (\ref{Lia_Z_LIT}) is present. It now follows that
$d(\delta^{2} S|_{\bf v})/dt \geq 0$.
\section{CONCLUSIONS}
In summary, when studying fluctuations in irreversible
thermodynamics using the formalism of Langevin or Fokker-Planck
equations, velocity is included as a variable. When making use of
the Einstein-Boltzmann relation to determine the exact form of the
fluctuation-dissipation relation the form of $\delta^{2} S$ where velocity
variation is allowed must be used. Although $S$ and $\delta S$ may be
written in forms that do not involve velocity, $\delta^{2} S$ does depend
on the velocity variation. If, as some authors do, $\delta^{2} S$ is taken
not to include velocity variations --- using what we have called
$\delta^{2} S|_{\bf v}$ --- then these velocity variations have to be
introduced by some other means, for example, by the introduction of the
$Z$ function. However, in this case an added postulate of the form
$P_{S} \sim \exp\left\{ \delta^{2} Z/2k_{B}\right\}$ has to be introduced.
Clearly, this is unnecessary since the usual Einstein-Boltzmann relation,
with the correct use of $\delta^{2} S$, that is, including velocity variations,
may be used without contradiction to complete the link between
thermodynamic and hydrodynamic fluctuations and the theory of
stochastic processes.

\vspace{0.9cm}

\noindent \textbf{Acknowledgements} We wish to thank Y. Oono and
M. L\'{o}pez de Haro for useful discussions. AJM wishes to thank
the Department of Physics at the Universidad Aut\'onoma del Estado
de Morelos for hospitality while this work was carried out. Financial
support from CONACYT-M\'exico under project number 40454 and from
PROMEP-M\'exico is gratefully acknowledged.

\end{document}